\documentclass[conference,a4paper]{IEEEtran}

\usepackage[utf8]{inputenc} 
\usepackage[T1]{fontenc}
\usepackage{url}              
\usepackage{cite}             

\usepackage[cmex10]{amsmath}  
\usepackage{mleftright}       
\mleftright                   

\usepackage{graphicx}         
\usepackage{booktabs}         

\usepackage{amssymb}
\usepackage{bm}
\def\t#1{\bm{\tilde{#1}}}
\def\ti#1{{\tilde{#1}}}
\def\bb#1{{\mathbb #1}}

\begin{document}

\title{Hubbard--Stratonovich Detector for Simple Trainable MIMO Signal Detection} 

\author{%
  \IEEEauthorblockN{Satoshi Takabe and Takashi Abe}
  \IEEEauthorblockA{%
  Tokyo Institute of Technology, Ookayama, Tokyo, 152-8550, Japan
}
}

%
%

\maketitle

\begin{abstract}
Massive multiple-input multiple-output (MIMO) is a key technology used in fifth-generation wireless communication networks and beyond.
 Recently, various MIMO signal detectors based on deep learning have been proposed. 
 Especially, deep unfolding (DU), which involves unrolling of an existing iterative algorithm and embedding of trainable parameters, has been applied with remarkable detection performance. 
 Although DU has a lesser number of trainable parameters than conventional deep neural networks, the computational complexities related to training and execution have been problematic because DU-based MIMO detectors usually utilize matrix inversion to improve their detection performance.
In this study, we attempted to construct a DU-based trainable MIMO detector with the simplest structure. 
The proposed detector based on the Hubbard--Stratonovich (HS) transformation and DU is called the trainable HS (THS) detector.  
It requires only $O(1)$ trainable parameters and its training and execution cost is $O(n^2)$ per iteration, where $n$ is the number of transmitting antennas.
Numerical results show that the detection performance of the THS detector is better than that of existing algorithms of the same complexity 
and close to that of a DU-based detector, which has higher training and execution costs than the THS detector.     
\end{abstract}

\section{Introduction}\label{sec_int}

Massive multiple-input multiple-output (mMIMO) has been considered a key technology in 
wireless communication systems such as the fifth-generation (5G) network and beyond~\cite{MUMIMO,Yang}. 
In mMIMO systems, tens or hundreds of antennas communicate simultaneously, resulting in high energy and spectral efficiency.
One challenge in mMIMO systems is \textit{overloading} in which the number of transmitting antennas is larger than that of receiving antennas.  
Signal detection in overloaded mMIMO systems requires linear regression of underdetermined systems, which further complicates the problem.
Recently, various MIMO signal detectors have been proposed to overcome the significant performance degradation of conventional detectors such as 
the minimum mean square error (MMSE) detector~\cite{MMSE}.
In particular, trainable detectors based on deep learning techniques show excellent detection performance~\cite{Bala}. 

Another challenge in mMIMO systems is related to ultra-reliable low-latency communication (URLLC)~\cite{Salh}, a key concept of post-5G networks. 
As MIMO systems become massive, the computational complexity of signal detection becomes crucial. 
Notably, a trainable MIMO detector has two types of computational complexity, i.e., \textit{training cost} and \textit{execution cost}.
The training cost relates to the scalability of the number of trainable parameters and the computational complexity of updating them in the training process. 
In a realistic scenario, trainable detectors should be learned online depending on the channel state information (CSI). 
This means that both training and execution costs are important for URLLC.  

The recent trainable MIMO detectors are mainly based on a deep learning technique called deep unfolding (DU)~\cite{LISTA,DU,DU2}. 
In DU, we first unroll the recursive structure of an iterative algorithm and embed some trainable parameters in each layer. 
These trainable parameters are updated by an optimizer to minimize a proper loss function whose value is determined by the supervised data and corresponding outputs of the unrolled algorithm.  
We can reduce the number of trainable parameters required by DU when compared with that required by conventional deep neural networks~\cite{TISTA2}.   
Numerous DU-based algorithms have been proposed for mMIMO systems, which include 
DetNet~\cite{DetNet}, trainable projected gradient (TPG) detector~\cite{TPG,TPG2}, and OAMP-net~\cite{He,He2}. 
Although these DU-based detectors show excellent performance, their training or execution cost continues to be high. 
For instance, in an mMIMO system with $n$ transmitting antennas, 
the training and execution cost of DetNet is $O(n^2)$ but it contains $O(n^2)$ parameters, whereas the
TPG detector and OAMP-net have only $O(1)$ parameters 
but require matrix inversion, resulting in $O(n^3)$ training and execution costs. 
These facts suggest the need for developing a trainable MIMO detector with low training and execution costs. 

The objective of this study was to develop a trainable MIMO detector with a simple structure. Accordingly, we propose an mMIMO detector with only $O(1)$ trainable parameters and computational cost of  
$O(n^2)$ for training and execution per iteration. 
To achieve our objective, we constructed a novel MIMO detector called Hubbard--Stratonovich (HS) detector inspired by the HS transformation~\cite{Hab,Stra}.
Then, by combining the HS detector with DU, we propose the trainable HS (THS) detector. 
We numerically analyzed its detection performance by comparing it with those of a scalable modified TPG detector and other conventional detectors. 

This paper is organized as follows. 
Section~\ref{sec_set} describes the MIMO system model.
In Sec.~\ref{sec_TPG}, we present an existing DU-based detector. 
Section~\ref{sec_HS} proposed HS and THS detectors.
Section~\ref{sec_cost} discusses the training and execution costs of trainable detectors.
We present the numerical results of the THS detector in Sec.~\ref{sec_res}, and  
discuss its superiority in Sec.~\ref{sec_dis}.
Section~\ref{sec_con} summarizes this paper.

 
\section{Model Setting}\label{sec_set}
In this section, we describe the channel model and introduce several definitions and notations.

The number of transmitting and receiving antennas is denoted by $n$ and $m$, respectively. If $m < n$ holds, the MIMO system is considered  overloaded.
For simplicity, we assume that the transmitter does not use precoding and that the receiver knows 
the perfect CSI, i.e., the channel matrix.

Let $\t{x} := [\ti{x}_1,\ti{x}_2,\dots,\ti{x}_n]^{T} \in \t{{\mathbb S}} ^{n}$ be 
a vector in which $\ti{x}_j$ ($j=1,\dots,n$) represents a transmitted symbol  from the $j$-th antenna.
The discrete set $\t{\bb{S}} \subset \bb{C}$ represents a signal constellation. 
Assuming a flat Rayleigh fading channel,  the received signal  $\t{y} := [\ti{y}_1,\ti{y}_2,\dots,\ti{y}_m]^{T} \in \bb{C}^{m}$ 
is given by 
$
  \t{y} = \t{H}\t{x} + \t{w}, 
$
where $\t{H} \in \bb{C}^{m \times n}$ is a channel matrix and 
$\t{w}\in \bb{C}^m$ is a complex additive white Gaussian noise vector   
with zero mean and covariance of $\sigma_{w}^2\bm I$.

This channel model can be rewritten in the real-number domain. 
The equivalent channel over $\mathbb{R}$ is given by
\begin{equation}
\bm y = \bm H\bm x + \bm w, \label{eq_rec}
\end{equation}
where 
   \begin{align} 
      \bm y &:= \begin{bmatrix}
        \Re(\t{y}) \\
        \Im(\t{y})
        \end{bmatrix} \in \bb{R}^{M},\ 
      \bm H := \begin{bmatrix} \label{eq:H_real}
        \Re(\t{H}) & - \Im(\t{H})\\
        \Im(\t{H}) &  \Re(\t{H})\\
      \end{bmatrix}, \\ 
      \bm{x} &:= \begin{bmatrix}
        \Re(\t{{x}}) \\
        \Im(\t{{x}})
        \end{bmatrix} \in \bb{S}^{N}, \ 
      \bm{w} := \begin{bmatrix}
        \Re(\t{{w}}) \\
        \Im(\t{{w}})
        \end{bmatrix}\in \bb{R}^{M},
    \end{align}
 and $(N,M) := (2n,2m)$.
 In the following, we consider the QPSK modulation, i.e., $\mathbb{S}=\{1,-1\}$, in the real-valued channel model (\ref{eq_rec}).

\section{TPG Detector}\label{sec_TPG}

Here, we describe the TPG detector~\cite{TPG,TPG2} as an example of the existing trainable MIMO detectors.

The MIMO signal detection is expressed as the following NP-hard optimization problem: 
\begin{equation}
\bm{\hat x} := \mathrm{argmin}_{\bm{{x}}\in\mathbb{S}^N} \frac{1}{2}\|\bm{{y}}-\bm{{H}{x}}\|_2^2. \label{eq_2}
\end{equation}
In particular, this is a linear regression in an underdetermined system if the system is overloaded, i.e., $M<N$. 

The TPG detector solves (\ref{eq_2}) using continuous relaxation and projected gradient descent. 
The update rule of the TPG detector is given by
\begin{align}
\bm r_t &= \bm s_t + \gamma_{t} \bm W(\bm y -\bm H \bm s_t ),\\
\bm s_{t+1} &= \tanh(\bm r_t/|\theta_{t}|),
\end{align}
where $\bm s_0=\bm 0$ and $\{\gamma_t,\theta_t\}_{t=0}^{T-1}$ is a set of trainable parameters. 
The $\tanh(\cdot)$ function is applied to each element.
The number of trainable parameters is omly $O(1)$, and the training cost is much lower than that of DetNet.   
In the original TPG detector, the matrix $\bm W$ is a linear MMSE (LMMSE)-like matrix $\bm H^T(\bm H \bm H^T + \alpha \bm I)^{-1}$ with a trainable parameter $\alpha$ because $\bm W = \bm H^T$ based on gradient descent shows poor detection performance~\cite{TPG2}. 
However, the LMMSE-like matrix requires matrix inversion, which has a computational cost of $O(n^3)$. 
Matrix inversion is required for each update in the training process, resulting in a training cost of $O(n^3)$.     
Another trainable detector called OAMP-net has the same problem.

\section{Hubbard--Stratonovich Detector}
\label{sec_HS}

Our goal is to propose a trainable MIMO signal detector without matrix inversions. 
Unlike the TPG detector based on continuous relaxation, we consider sampling a solution of (\ref{eq_2}) using the Boltzmann distribution given by 
 \begin{equation}
P(\bm x;\beta):= \frac{1}{Z}\exp\left( -\frac{\beta\lambda}{2}\|\bm{{y}}-\bm{{H}{x}}\|_2^2\right) , \label{eq_B}
\end{equation}
where $\beta(>0)$ is called the inverse temperature, $\lambda(>0)$ is a scaling parameter, and $Z$ is a normalization constant called the partition function.
 An optimal solution of (\ref{eq_2}) is obtained in the $\beta\to\infty$ limit, but calculating the probability is difficult because of $Z$.

Here, we attempt to sample $\bm x$ from (\ref{eq_B}) by using an efficient method proposed by Ohzeki~\cite{OH}. 
The key step is the use of an identity known as the HS transformation in statistical physics~\cite{Hab,Stra}: for $a>0$, 
\begin{equation}
\exp\left(-\frac{ax^2}{2}\right)=\int \frac{1}{\sqrt{2\pi a}}\exp\left(-\frac{z^2}{2a}-ixz\right)dz. \label{eq_HS}
 \end{equation}
 Recently, the HS transformation is also applied to probabilistic inference in machine learning~\cite{Koe}. 
 Then, the partition function $Z$ is rewritten as follows: 
 \begin{align}
Z &:= \sum_{\bm{x}\in\mathbb{S}^{N}} \exp\left( -\frac{\beta\lambda}{2}\|\bm{{y}}-\bm{{H}{x}}\|_2^2\right) \nonumber\\
&= \int \left(\prod_{j=1}^{M}\frac{dz_j}{\sqrt{2\pi \beta\lambda}}\right)  
\sum_{\bm{x}\in\mathbb{S}^{N}} \exp\left( -\frac{\|\bm z\|_2^2}{2\beta\lambda} - i \bm z^T (\bm{Hx}-\bm{y})\right) \nonumber\\
&\propto  \int d\bm v \sum_{\bm{x}\in\mathbb{S}^{N}}\exp\left( \frac{\beta \|\bm v\|_2^2}{2\lambda} + \beta \bm v^T (\bm{Hx}-\bm{y}) \right)   \nonumber\\
&:=\int d\bm v e^{-\beta H(\bm v)}, \label{eq_z}
\end{align}
 where $\bm z = (z_1,\dots,z_M)^T$, $\bm v = (v_1,\dots,v_M)^T$,  and $z_j = i \beta v_j$ ($j=1,\dots, M$).

 If $\beta$ is sufficiently large, the integral can be estimated by the saddle-point method, i.e.,  
 $Z\propto e^{-\beta H(\bm{\hat v})} $, where $\bm{\hat v}$ is a saddle point of the following function:
\begin{align}
H(\bm v)& = -\frac{\|\bm v\|_2^2}{2\lambda} + \bm v^T\bm{y} - \frac{1}{\beta}\ln Z(\bm v),  \label{eq_hv1} \\
Z(\bm v) &:= \sum_{\bm{x}\in\mathbb{S}^{N}} e^{\beta \bm v^T \bm H \bm x}. \label{eq_hv2}
\end{align}
We can also estimate the expectation of $\bm x$ by $\bm{\hat v}$ because, using the same transformation, we have
\begin{align}
\sum_{\bm{x}\in\mathbb{S}^{N}} \bm xP(\bm x;\beta)& := \frac{1}{Z}\int d\bm v \!\sum_{\bm{x}\in\mathbb{S}^{N}} \bm x \exp\left( \frac{\beta \|\bm v\|_2^2}{2\lambda}\! + \!\beta \bm v^T (\bm{Hx}\!-\!\bm{y}) \right) \nonumber\\
&\simeq  \frac{\sum_{\bm{x}\in\mathbb{S}^{N}} \bm x e^{\beta \bm{\hat v}^T \bm{Hx}} }{\sum_{\bm{x}\in\mathbb{S}^{N}}  e^{\beta \bm{\hat v}^T \bm{Hx}}}. \label{eq_exx}
\end{align}
The expectation is equal to $\bm{\hat x}$ if the optimization problem (\ref{eq_2}) has a unique optimal solution.

Based on  gradient descent, a simple update rule to estimate $\bm{\hat v}$ is obtained~\cite{OH}. 
The derivative of (\ref{eq_hv1}) contains
the expectation of $\bm x$ defined as
\begin{equation} 
\left<\bm x\right>_{\bm v}:= \frac{\sum_{\bm{x}\in\mathbb{S}^{N}} \bm  x e^{\beta \bm v^T \bm H \bm x}}{\sum_{\bm{x}\in\mathbb{S}^{N}} e^{\beta \bm v^T \bm H \bm x}}. \label{eq_app3}
\end{equation}
This is equivalent to (\ref{eq_exx}) when  $\bm v=\bm{\hat{v}}$. 
For $i=1,\dots,N$, the $i$-th element of (\ref{eq_app3}) is easily calculated because $\{x_i\}$ become independent random variable in this problem setting.  
We find 
\begin{align}
\frac{\sum_{\bm{x}\in\mathbb{S}^{N}} x_i e^{\beta \bm v^T \bm H \bm x}}{\sum_{\bm{x}\in\mathbb{S}^{N}} e^{\beta \bm v^T \bm H \bm x}} &=  \frac{\sum_{x_i \in\mathbb{S}} x_i e^{\beta (\bm H^T\bm v)_i x_i}}{\sum_{x_i\in\mathbb{S}} e^{\beta (\bm H^T\bm v)_i x_i}} \nonumber\\
& = \tanh(\beta (\bm H^T\bm v)_i ). \label{eq_app4}
\end{align}

Introducing $\bm u:= \bm H^T \bm v$, we have the update rule given by 
\begin{align}
\bm u_{t+1} &= \left(1+\frac{\eta}{\lambda}\right)\bm u_t + \eta \bm H^T (\bm y -\bm H \bm s_t ), \label{eq_HS1}\\ 
\bm s_{t+1} &= \tanh(\beta \bm u_{t+1}),\label{eq_HS2}
\end{align}
where $\bm u_0\!=\!\bm s_0\!=\!\bm 0$ and $\eta$ is a step-size parameter. 
The parameter $\lambda$ controls the strength of the quadratic term of (\ref{eq_hv1}). 
We call this the Hubbard--Stratonovich (HS) detector.
Note that if $\bm u_t=\bm H \bm{\hat v}$ holds, $\bm s_t$ is equivalent to (\ref{eq_exx}) and possibly $\bm{\hat x}$. 
Thus, the HS detector can be considered the expectation-maximization (EM) algorithm~\cite{EM} that searches 
 $\bm{\hat v}$ and $\bm{\hat x}$ simultaneously, where (\ref{eq_HS1}) and (\ref{eq_HS2}) correspond to the maximization step and expectation step, respectively.    
It is also noteworthy that the difficulty in solving (\ref{eq_2}) by the HS detector is related to finding $\bm{\hat v}$ because the function $H(\bm v)$ is non-convex at least in the $\beta\to\infty$ limit. 
  
Finally, DU is applied to the HS detector to train some internal parameters. 
The trainable MIMO detector, called the THS detector, is defined as 
\begin{align}
\bm u_{t+1} &= \zeta_{t} \bm u_t + \eta_{t} \bm H^T(\bm y -\bm H \bm s_t ), \label{eq_THS1}\\ 
\bm s_{t+1} &= \tanh(\beta_{t} \bm u_{t+1}),\label{eq_THS2}
\end{align}
 where $\bm u_0\!=\!\bm s_0\!=\!\bm 0$. 
We introduce $\{\beta_t, \eta_t,\zeta_t\}_{t=0}^{T-1}$ as trainable parameters, which affect the detection performance. 
 The inverse temperature $\beta_t$ is tuned because a large $\beta_t$ leads to fast convergence to an undesired local minima, whereas $\beta_t$ should be infinitely large as described above. 
 The step-size $\eta_t$ controls the convergence speed, and $\zeta_t$ is related to $\lambda$ in (\ref{eq_HS1}). 
       
\section{Training and Execution Costs}
\label{sec_cost}

Here, we discuss the training and execution costs of the THS detector.
It is obvious that the THS detector contains only three trainable parameters in each iteration (or layer), and that 
the computational complexity remains limited to $O(n^2)$ owing to the absence of matrix inversion.

In Tab.~\ref{tab_1}, we summarize the number of trainable parameters and the training and execution costs of the trainable MIMO detectors per iteration (or layer). 
Note that the training cost indicates the computational complexity required to update the trainable parameters in the training process whereas the execution cost is the computational cost for executing the learned algorithm. 
In particular, the TPG detector requires matrix inversion once during the execution, which is omitted from the execution cost per iteration in Tab.~\ref{tab_1}. 
However, it requires a matrix inversion every time a trainable parameter $\alpha$ is updated in the training process, indicating that the training cost per update is $O(n^3)$.  
We find that the THS detector is advantageous in terms of the number of trainable parameters when compared with DetNet. 
In addition, the THS detector is more computationally efficient than the TPG detector and OAMP-net with matrix inversions.
Overall, the THS detector has the simplest structure in terms of the number of trainable parameters and the training and execution costs,  
implying fast training and execution for mMIMO systems.

 \begin{table}[t]
   \centering
  \caption{Number of trainable parameters and 
computational costs for training and execution per iteration. }
   \label{tab_1}
   \begin{IEEEeqnarraybox}[\IEEEeqnarraystrutmode%
     ]{l"l"l"l"l}
     \toprule
      & $THS$ & $DetNet$ & $TPG$ & $OAMP-net$  \\
     \midrule
     $\# of parameters$ & O(1) & O(mn)  & O(1) & O(1) \\      
     $Training cost$ &  O(n^2) & O(n^2) & O(n^3) & O(n^3) \\
     $Execution cost$  & O(n^2) &  O(n^2) & O(n^2)^{\mbox{\dag}} & O(n^3) \\
     \bottomrule
   \end{IEEEeqnarraybox}\\
   $^{\mbox{\dag}}$ The TPG detector requires matrix inversion once before its iterations.
 \end{table}

\section{Numerical Results}
\label{sec_res}

In this section, we present the numerical results of the signal detection performance of the proposed THS detector.

\subsection{Simulation Settings}

We first describe the simulation settings. 
In the MIMO system (\ref{eq_rec}), the signal-to-noise ratio (SNR) is given by $\mathrm{SNR}=10\log_{10}\frac{n}{\sigma_w^2}$. We estimate the bit error rate (BER) as the indicator for signal detection performance. 
All real-valued detected signals are thresholded to $\mathbb{S}=\{1,-1\}$ using the sign function $\mathrm{sign}(*)$ for calculating the BER. 

The THS detector was implemented using PyTorch 1.13.1~\cite{PyTorch}.
Its training is based on supervised learning in which pairs of transmitted and received signals $\{(\bm x,\bm y)\}$ are randomly generated. These training data are fed to the THS detector as a mini-batch.  
In each parameter update, 1000 mini-batches of size 2000 are fed, and the parameters are updated by an Adam optimizer~\cite{Adam} with a learning rate of $2\times 10^{-4}$ to minimize the MSE loss function. 
A channel matrix $\bm H$ is generated for each mini-batch.
We employed incremental training to ensure the stability of the training process (see~\cite{TPG2} for details).
The number of iterations was set to $T=30$, and the initial values of the trainable parameters were $\eta_t=0.01$ and $\beta_t=\zeta_t=1$. 

We also examined the MMSE detector, enhanced reactive tabu search (ERTS)~\cite{ERTS}, HS detector, TPG detector, and its variant as the baseline algorithms. 
MMSE and ERTS are conventional MIMO detectors. The computational complexity of the MMSE detector is $O(n^3)$.  
ERTS searches a transmitted signal from the random initial points. Its computational cost is $O(Kn^2)$, 
where $K$ is the number of iterations. We set $K=500$ in the simulations.    
The HS detector is executed with fixed step-size parameters $\eta=0.1$ and $\lambda=1$. 
The TPG detector was trained and executed under the same conditions as in~\cite{TPG2}.
In addition, we examined the TPG detector with $\bm W=\bm H^T$, called the scalable TPG detector; it has the same training and computational costs as the THS detector.  
The number of iterations of the HS, TPG, and scalable TPG detectors was set to $T=30$, as in the case of the THS detector.

\subsection{Numerical Results}

We present the detection performance of overloaded mMIMO systems.

\begin{figure}[!t]
\centering
\includegraphics[width=0.94\linewidth]{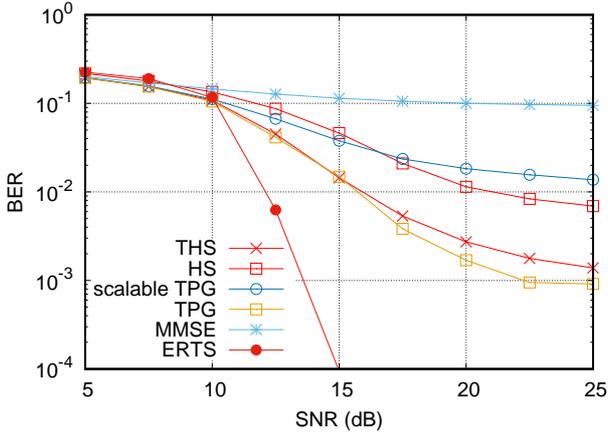}
\caption{BER performance for $(n,m)=(50,32)$. The number of iterations of (T)HS and (scalable) TPG detectors is $T=30$ and that of ERTS is $K=500$.
}\label{zu_s1}
\end{figure}

\begin{figure}[!t]
\centering
\includegraphics[width=0.94\linewidth]{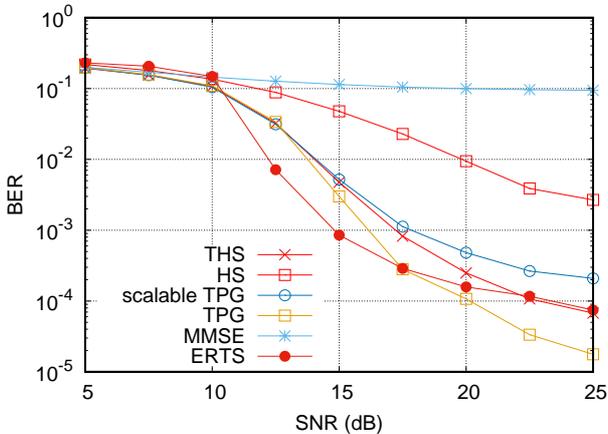}
\caption{BER performance for $(n,m)=(100,64)$. 
}\label{zu_s2}
\end{figure}

Figure~\ref{zu_s1} depicts the detection performance when $(n,m)=(50,32)$. 
In this case, ERTS shows the best performance but the number of iterations $K=500$ is much larger than that of the (T)HS and TPG detectors. 
Interestingly, the HS detector without learning shows better performance than the scalable TPG detector in the high-SNR regime. 
The detection performance of the THS detector was superior to that of the HS detector, indicating that DU successfully improves the detection performance; when $\mathrm{SNR}=20$ dB, the BER of the THS detector was approximately $2.7\times 10^{-3}$ whereas 
those of the HS and scalable TPG detectors were $1.1\times 10^{-2}$ and $1.8\times 10^{-2}$, respectively. 
In addition, the detection performance of the THS detector was close to that of the TPG detector but the training cost was lower. 
The SNR gap between the THS and TPG detectors was approximately $2.5$ dB at $\mathrm{BER}=2.0\times 10^{-3}$.
  
Figure~\ref{zu_s2} shows the detection performance when $(n,m)=(100,64)$. 
In this case, the performance of the HS detector degrades significantly when compared with that of the THS detector, although it is superior to that of the MMSE detector. This is because the convergence speed of the HS detector becomes slower as the antenna size increases. 
Similarly, the performance of ERTS degrades because of the curse of dimensionality in searching for a good estimation. 
The THS detector showed better performance than the scalable TPG detector but poorer performance than the TPG detector; 
further, it achieved a gain of approximately $2.5$ dB when compared with the scalable TPG detector at $\mathrm{BER}=2.0\times 10^{-4}$. In the high-SNR regime, the performance of the THS detector was fairly close to that of ERTS. 

\begin{figure}[!t]
\centering
\includegraphics[width=0.94\linewidth]{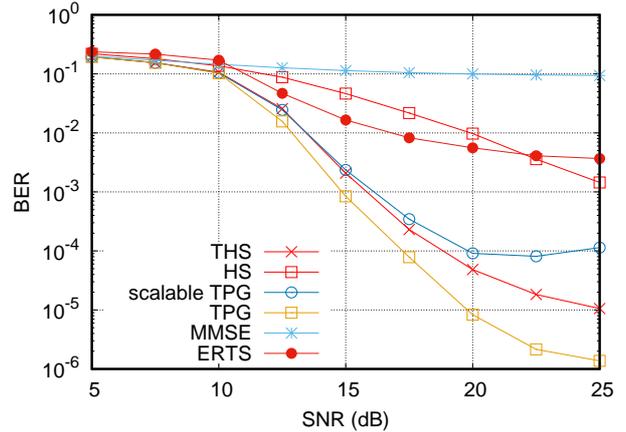}
\caption{BER performance for $(n,m)=(150,96)$. 
}\label{zu_s3}
\end{figure}

Figure~\ref{zu_s3} shows the detection performance when $(n,m)=(150,96)$. 
We find that the detection performances of the ERTS and HS detectors are significantly degraded. 
In addition, the scalable TPG detector shows an error floor around $\mathrm{BER}=10^{-4}$ in the high-SNR regime.
 However, the THS detector exhibits better detection performance, especially in the high-SNR regime, whereas its SNR gap from the TPG detector increases. 
 When $\mathrm{SNR}=20$ dB, the BER of the THS detector is approximately $1.1\times 10^{-5}$ whereas that of the scalable TPG detector is approximately $1.1\times 10^{-4}$.

We can conclude that the THS detector successfully detects the transmitted signals in overloaded mMIMO systems.
Unlike the HS, scalable TPG, and ERTS detectors, the proposed THS detector shows excellent performance regardless of the number of antennas. 
Moreover, it achieves a performance close to that of the TPG detector requiring matrix inversion.

\begin{figure}[!t]
\centering
\includegraphics[width=0.94\linewidth]{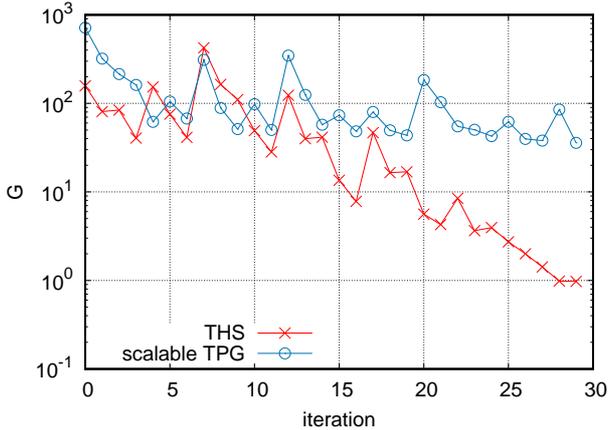}
\caption{Average amplitude of gradient $G$ of the THS and scalable TPG detectors over $10^4$ signals for a noiseless MIMO system with $(n,m)=(50,32)$.  The BER performances of the THS and scalable TPG detectors are $1.1\times 10^{-3}$ and  $4.8\times 10^{-3}$, respectively.
}\label{zu_s4}
\end{figure}

\section{Discussion: Difference between THS and Scalable TPG Detectors}
\label{sec_dis}

In the previous section, the numerical results showed that the THS detector exhibited better detection performance than the scalable TPG detector. In this section, we discuss the reason for the successful signal detection of the THS detector.
 
First, we present the update rules of the two detectors. 
The update rule of the scalable TPG detector is given by 
\begin{equation}
\bm r_{t+1} = \tanh(\beta_t \bm r_t) + \eta_{t+1} \bm H^T[\bm y -\bm H \tanh(\beta_t \bm r_t) ], \label{eq_sc}
\end{equation}
where we set $\eta_t=\gamma_t$ and $\beta_t=|\theta_t|^{-1}$. 
In this update rule, $\bm r_t$ is a search point of the problem (\ref{eq_2}), neglecting the constraint $\bm x\in\mathbb{S}^N$.
To investigate a fixed point of (\ref{eq_sc}), let us consider a noiseless case.  
Then,  the transmitted signal $\bm{x}_0$ is the exact solution of (\ref{eq_2}) and 
 should be a fixed point of (\ref{eq_sc}) because $\bm y = \bm H \bm x_0$ holds. 
 However, it is easily found that $\bm r^\ast $ satisfying $\bm x_0=\tanh(\beta_t \bm r^\ast)$ cannot be a fixed point unless $\beta_t\to \infty$ owing to the soft projection function $\tanh(\cdot)$. 
Recalling that the TPG detector with $\beta\to\infty$ usually fails signal detection, it is difficult to find a fixed point in $\mathbb S$ for the TPG detector with finite $\beta_t$. 

On the other hand, the THS detector is defined as 
\begin{equation}
\bm u_{t+1} =  \bm u_t + \eta_{t+1} \bm H^T[\bm y -\bm H \tanh(\beta_t \bm u_t) ], \label{eq_ths}
\end{equation}
where we assume $\zeta_t=1$ (or $\lambda\to\infty$). 
The difference between the THS (\ref{eq_ths}) and scalable TPG (\ref{eq_sc}) detectors lies only in the first term. 
However, as described in Sec.~\ref{sec_HS}, $\bm u_t$ relates to $\bm v$ of the HS transformation, and is regarded as local fields of $\bm x$, i.e., $P(x_i)\propto e^{\beta_t u_{ti}x_i}$, where $u_{ti}$ is the $i$-th element of $\bm u_t$.     
In the noiseless case, the THS detector has a fixed point $\bm u^\ast$ corresponding to $\bm x_0=\tanh(\beta_t \bm u^\ast)$ for any $\beta_t>0$, unlike the scalable TPG detector. 

Additionally, we numerically studied these detectors in a noiseless MIMO system with $(n,m)=(50,32)$. 
In this case, the BERs of the THS and scalable TPG detectors were $1.1\times 10^{-3}$ and  $4.8\times 10^{-3}$, respectively. The trained values of $\beta_t$ of the scalable TPG took a range from $0.24$ to $2.65$, indicating that the transmitted signal is not a fixed point of (\ref{eq_sc}).

Figure~\ref{zu_s4} shows the average amplitude of the gradient defined by 
$G:= {N}^{-1} \| \bm H^T(\bm y -\bm H \bm s_t )\|_2$. 
This value should be zero if $\bm s_t=\bm x_0$.  
We find that the value for the THS detector gradually decreases as the number of iterations increases,
wheras that for the TPG detector remains large.

\begin{figure}[!t]
\centering
\includegraphics[width=0.98\linewidth]{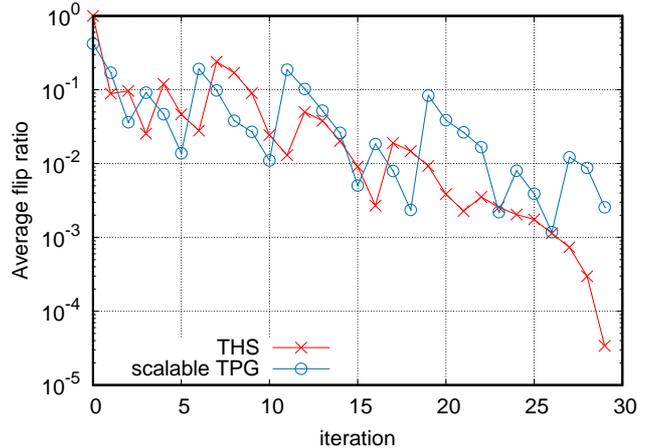}
\caption{Average bit-flip ratio of the THS and scalable TPG detectors over $10^4$ signals for a noiseless MIMO system with $(n,m)=(50,32)$. 
}\label{zu_s5}
\end{figure}

In Fig.~\ref{zu_s5}, we depict the average bit-flip ratio, which denotes the fraction of bit flips from $\mathrm{sign}(\bm s_{t})$ to $\mathrm{sign}(\bm s_{t+1})$.
We find that the bit-flip ratio of the THS detector drops as $G$ decreases whereas that of the TPG detector remains relatively large. 
These differences suggest that the search point of the THS detector approaches a desired fixed point but that of the scalable TPG detector is far from the fixed point. 
This numerical stability is another possible reason for the difference in the detection performance in the high-SNR regime.


\section{Concluding Remarks}
\label{sec_con}

In this paper, we proposed a novel mMIMO detector 
 inspired by HS transformation and DU. 
The proposed detector, called the THS detector, has a simple update rule with $O(1)$ trainable parameters per iteration. 
In addition, its training and execution cost is $O(n^2)$ per iteration, which is the least among the existing trainable MIMO detectors, making it suitable for URLLC. 
Numerical results showed that the THS detector performs better than the scalable TPG detector, which has the same computational costs. In addition, the performance is fairly close to that of the TPG detector but the computational costs are lower. 
We also discussed the successful detection of the THS detector, and showed that it is more stable than the scalable TPG detector, as its search point is close to a fixed point in a noiseless case. 
In our future study, we intend to investigate the THS detector for higher-order modulations and coded modulations.  The application of the Chebyshev steps~\cite{ch,ch2} will be beneficial in reducing the number of trainable parameters. 
A theoretical analysis of the HS and THS detector is also an interesting research topic.    

\section*{Acknowledgement}
We thank Masayuki Ohzeki for discussions on~\cite{OH}. 
ST also thanks Takashi Takahashi, Jun Takahashi, Yoshihiko Nishikawa, and Harukuni Ikeda for fruitful discussions.  
This work was partly supported by  JSPS KAKENHI Grant Numbers 22H00514, 22K17964, and  19K14613.


 \IEEEtriggeratref{11}

\end{document}